\documentclass[prd,aps,twocolumn,amsmath,amssymb,floatfix,nofootinbib,superscriptaddress]{revtex4}

\usepackage{t1enc}
\usepackage{amsmath,amssymb,amsfonts}
\usepackage{gensymb}
\usepackage{graphicx}
\usepackage{float}
\usepackage{physics}
\usepackage{booktabs}
\usepackage{slashed}

\newcommand{\drm}{\ensuremath{\mathrm{d}}}

\newcommand{\f}[2]{\frac{#1}{#2}}

\newcommand{\Oc}{{\cal O}}

\renewcommand{\Im}{{\rm Im}}

\newcommand{\Dmeas}{\mathcal{D}}

\begin{document}

\title{Fighting the sign problem in a chiral random matrix model with contour deformations}

\author{Matteo Giordano}
\affiliation{ELTE E\"otv\"os Lor\'and University, Institute for Theoretical Physics, P\'azm\'any P\'eter s\'et\'any 1/A, H-1117, Budapest, Hungary}

\author{Attila P\'asztor}
\affiliation{ELTE E\"otv\"os Lor\'and University, Institute for Theoretical Physics, P\'azm\'any P\'eter s\'et\'any 1/A, H-1117, Budapest, Hungary}

\author{D\'avid Peszny\'ak}
\affiliation{ELTE E\"otv\"os Lor\'and University, Institute for Theoretical Physics, P\'azm\'any P\'eter s\'et\'any 1/A, H-1117, Budapest, Hungary}

\author{Zolt\'an Tulip\'ant}
\affiliation{ELTE E\"otv\"os Lor\'and University, Institute for Theoretical Physics, P\'azm\'any P\'eter s\'et\'any 1/A, H-1117, Budapest, Hungary}

\begin{abstract}
We studied integration contour deformations in the chiral random matrix theory of Stephanov~\cite{Stephanov:1996ki} with the goal of alleviating the finite-density sign problem. We considered simple ansätze for the deformed integration
contours, and optimized their parameters. We find 
that optimization of a single parameter manages to considerably improve on the severity of the sign problem. We show numerical evidence that the improvement achieved is exponential in the degrees of freedom of the system, i.e., the size of the random matrix. We also compare the optimization method with 
contour deformations coming from the holomorphic flow
equations.
\end{abstract}

\maketitle

\section{Introduction}

Euclidean quantum field theories at non-zero particle density 
(or chemical potential) generally suffer from a complex 
action problem: the weights in the path integral representation 
are complex, 
and thus cannot be interpreted as a joint probability density function on the space of field configurations (up to a proportionality factor).
This prevents the use of importance sampling methods 
for the direct simulation of these theories. In QCD, this 
complex action problem severely hampers first-principles studies 
of dense matter in the core of neutron stars, neutron star mergers, 
core collapse supernovae, as well as in heavy ion collisions at 
certain collision energies. 

In the presence of a complex action problem one can still (in principle) simulate a modified theory with real and positive weights, and then use reweighting methods to calculate observables in the theory of interest.
If the target theory has field variables $\phi$, path integral weights $w_t(\phi)$, and partition function
$Z_t = \int\Dmeas \phi \ w_t(\phi)$, and the simulated theory has 
the same field variables, but different 
-- real and positive -- path integral weights $w_s(\phi)$ and partition function $Z_s = \int\Dmeas \phi \ w_s(\phi)$,
we can obtain expectation values in the target theory via the formula 
\begin{equation}
  \label{eq:reweight}
  \left\langle \Oc \right\rangle_t = \frac{\left\langle \frac{w_t}{w_s}
      \Oc \right\rangle_s }{\left\langle
      \frac{w_t}{w_s}\right\rangle_s}\,,\quad
  \langle \Oc \rangle_x = \f{1}{Z_x} \int\Dmeas \phi \ w_x(\phi) \Oc(\phi)\,,
\end{equation}
where $x$ may stand for $t$ or $s$ and $\Oc(\phi)$ is some physical observable
of interest. The denominator in Eq.~\eqref{eq:reweight} gives the ratio of 
the partition functions in the target and simulated theories, i.e.,
\begin{equation}
  \label{eq:reweightZ}
    \left\langle \frac{w_t}{w_s} \right\rangle_s = \frac{Z_t}{Z_s}\rm{.}
\end{equation}
This ratio is typically exponentially small in the physical volume, with 
the exponent given by the free energy difference between the target and 
simulated theories. This ratio is also a rough measure of the numerical 
difficulty of a given reweighting scheme, with a given simulated and target 
theory. In order for reweighting to be effective, one wants 
the target and simulated theories to be as close to each other as 
possible. Ideally, one should find a simulated theory with $Z_s \approx Z_t$. 

Two simple choices of a simulated theory are the phase-quenched (PQ) theory, with simulated weights proportional to 
\begin{equation}
    w_s^{\mathrm{PQ}} \equiv \left| w_t(\phi)\right|\rm{,} 
\end{equation}    
or -- assuming that the partition function $Z_t$ is real -- 
the sign-quenched (SQ) theory, with simulated weights proportional to
\begin{equation}
    w_s^{\mathrm{SQ}} \equiv \left| \operatorname{Re} w_t(\phi) \right|\rm{.}
\end{equation}    
For the first case (phase reweighting) the reweighting factors $w_t / w_s^{\mathrm{PQ}} \equiv e^{i \theta}$ are pure phases.
For the second case (sign reweighting) the reweighting factors are $w_t / w_s^{\mathrm{SQ}} = e^{i \theta}/ \left| \cos \theta \right|$.
For certain observables, such as manifestly real observables or observables with a conjugation ($\phi \to \overline{\phi}$) symmetry, one can substitute $w_t/w_s^{\mathrm{PQ}}$ with $\cos\theta$ and $w_t/w_s^{\mathrm{SQ}}$ with a pure sign $\cos\theta/\left|\cos\theta\right|$. For phase or sign reweighting, we can then say that the complex action problem becomes a sign problem: the cancellations between contributions with different signs of $cos\theta$ lead to a small $\frac{Z_t}{Z_s}$ ratio, and in turn to small signal-to-noise ratios in the expectation values of observables.

The sign-quenched ensemble always has a less severe sign problem, due to the 
inequality $Z_t < Z_s^{\mathrm{SQ}} < Z_s^{\mathrm{PQ}}$, which is a consequence of $\cos \theta \leq |\cos \theta| \leq 1$. However, in the limit of a severe sign problem  -- i.e., as the distribution of the argument $\theta$ tends to to a uniform distribution on $[-\pi,\pi)$ -- the severity of the sign problem for these two reweighting schemes
only differs by a constant factor~\cite{Borsanyi:2021hbk}, given by $\left( Z_s^{\mathrm{PQ}} /Z_s^{\mathrm{SQ}} \right)^2 \to \left( \pi / 2 \right)^2$.  

In QCD and in other (more or less) QCD-like models, describing the interactions of several ``flavors'' of fermions, the path integral weights can be written schematically as 
\begin{equation}
    w_t(\phi) = \det M_1(\phi,\mu_1) \ldots \det M_{N_f}(\phi,\mu_{N_f}) e^{-S_B(\phi)}\rm{,}
\end{equation}
where the fields $\phi$ are real bosonic variables and $S_B$ is the corresponding bosonic part of the action,  $N_f$ is the number of fermion flavors in the model, $\det M_k $ is the fermionic determinant of the $k$th flavor and $\mu_k$ is the corresponding chemical potential,  for $k = 1, \ldots , N_f$.  The source of the sign problem is the fermionic determinant, which at non-zero $\mu$ is generally a complex number. Moreover, an important feature of the sign problem in QCD  and QCD-like theories is that it tends to get much worse in the ranges of $\mu$ where zeros of the determinant in the complex $\mu$ plane become dense~\cite{Nagata:2021ugx}.

Nonetheless, reweighting from the phase- and 
sign-quenched theories is 
starting to become 
feasible even in full QCD~\cite{Giordano:2020roi,Borsanyi:2021hbk}, 
which has recently 
led to the calculation of the equation of state of a 
hot-and-dense quark-gluon plasma in the region of chemical potentials 
covered by the RHIC Beam Energy Scan~\cite{Borsanyi:2022soo}. However, 
the range of practical applicability of such an approach is limited both
in volume and chemical potential by the smallness of the 
ratio $Z_t/Z_s$. 
Lacking a solution of the sign problem, it is then desirable to develop methods that at least alleviate it, to extend the range of parameters that reweighting methods can practically reach.

One possible route to do this is the use of contour 
deformations in the path integral (see Ref.~\cite{Alexandru:2020wrj} for a 
recent review). If the path integral weights 
$w_t(\phi)$ are holomorphic functions of the field variables,\footnote{A 
notable exception is lattice QCD with rooted staggered 
fermions~\cite{Golterman:2006rw,Giordano:2019gev}.} 
the multivariate Cauchy theorem guarantees that complexified integration
manifolds in the same homology class as the original one yield the same 
partition function. However, the phase- and sign-quenched integrands are 
not holomorphic, and therefore the phase- and sign-quenched partition functions 
are not invariant under such deformations. It may then 
be possible to bring the ratios $Z_t/Z_s$ closer to unity, thus 
making reweighting more effective. 

There are different ways to deform integration contours. Historically, 
methods based on Lefschetz thimbles appeared 
first~\cite{Cristoforetti:2012su, Cristoforetti:2013wha, Alexandru:2015sua, Fukuma:2020fez, Alexandru:2020wrj, DiRenzo:2020cgp,DiRenzo:2021kcw}.
Lefschetz thimbles are the disjoint components of the integration contour defined by requiring that the imaginary part of the classical action is
constant in each component.
The thimble structure of theories with a 
fermionic determinant is usually quite complicated~\cite{Kanazawa:2014qma, Tanizaki:2015rda, DiRenzo:2017igr, Zambello:2018ibq, Ulybyshev:2019fte}. Simple toy models reveal the following features:
i) cancellations between competing thimbles are very important for getting the correct results, and ii) the thimbles themselves are not smooth at the zeros of the fermionic determinant. Thus, the use of thimbles might be impractical for such theories.
However, Lefschetz thimbles are, in general, not the 
numerically optimal integration 
contours~\cite{Lawrence:2018mve}, i.e., they are not necessarily the contours with the largest $Z_t/Z_s$, so there is no need to concentrate solely on them.

A second class of methods is based on numerical optimization. 
The main idea here is to parametrize the integration manifold 
by a finite number of parameters, 
which are then optimized to make the sign problem as mild 
as possible. Such methods were applied to a 
one-dimensional integral~\cite{Mori:2017pne}, the 
0+1D scalar theory~\cite{Bursa:2018ykf}, the 
0+1D Polyakov-improved Nambu-Jona-Lasinio model~\cite{Kashiwa:2018vxr}, 0+1D QCD~\cite{Mori:2019tux}, 
1+1D scalar field theory~\cite{Mori:2017nwj}, the 
1+1D Thirring model~\cite{Alexandru:2018fqp}, the 2+1D 
Thirring model~\cite{Alexandru:2018ddf}, Bose gases of several dimensions~\cite{Bursa:2021org}, 1+1D U(1) gauge 
theory with a complex coupling constant~\cite{Kashiwa:2020brj} 
and the 2+1D XY model at finite density~\cite{Giordano:2022miv}.
Here, we apply contour optimization methods 
to a fermionic toy model 
that
shares relevant technical features with finite chemical potential QCD: the 
chiral random matrix model proposed by Stephanov
in Ref.~\cite{Stephanov:1996ki}.

Since it is an exactly solvable model with a sign problem, the Stephanov model is a very useful testbed for methods aimed at solving or alleviating the sign problem. This model has been studied with the
complex Langevin approach
~\cite{Parisi:1983mgm, Aarts:2009uq,Seiler:2012wz}, 
which fails 
for this particular model~\cite{Bloch:2017sex} even with 
the introduction of gauge cooling~\cite{Seiler:2012wz}.
There are also preliminary results for this model with the 
tempered Lefschetz thimble method~\cite{Fukuma:2022yhy} which is
based on parallel tempering~\cite{PhysRevLett.57.2607} in 
the flow time of the 
holomorphic flow~\cite{Alexandru:2015sua,Fukuma:2017fjq}.
This method -- similarly to other flow-based methods --- produces a weaker sign problem, albeit at the cost of substantially 
increasing the per-configuration-cost of generating 
the ensemble compared to ordinary phase reweighting.

In this paper we study the Stephanov model with optimization methods. There are, roughly speaking, two approaches to such an optimization: one can look for the optimum using either a very general ansatz with a large number of parameters, or a very specific ansatz tailored for the model at hand, and with a small number of parameters. The first approach has clearly the potential to find a good optimum, e.g., using machine learning techniques, but it also has some disadvantages. In fact, for such a general approach  the number of optimization parameters has to be increased as one increases the number of degrees of freedom of the system. This means that the cost of finding good contours might turn out to be prohibitive, similarly to what happens with methods based on Lefschetz thimbles. In this exploratory study we follow the second, ad hoc approach, and optimize ansätze with only few parameters.
Moreover, the number of these parameters is kept independent of the number of degrees of freedom of the system. We can then be sure that the optimization itself is numerically cheap, and that the per-configuration cost of generating the ensembles is essentially as low as on the original contours. Obviously, the drawback of this approach is that to write down an ansatz with only a few parameters that produces a substantial improvement in the severity of the sign problem, some physical or mathematical insight is needed. 

For the toy model studied in this paper, the 
insight required to use the ad hoc approach is 
available, and so we can write down appropriate 
ansätze. We will then show that a quite cheap numerical 
optimization procedure leads one to contours with a reduced sign problem. 
We will also present numerical evidence that the reduction in the severity of 
the sign problem is exponential: while the sign problem 
on the optimized contours is still exponential in the number of degrees of 
freedom, the corresponding exponent is reduced. This conclusion is similar to 
what some of us have shown in Ref.~\cite{Giordano:2022miv} for a 
purely bosonic model (the 2+1 dimensional XY model 
at non-zero chemical potential). Notably, such an 
exponential reduction can be achieved without changing the 
number of optimization parameters with the system size.

In this work we will only consider phase-quenched simulations, for simplicity. Similar arguments and methods should, however, also apply to the sign-quenched case~\cite{Giordano:2022miv}.

The plan of the paper is the following:
In Section~\ref{sect:model} we introduce the model
discussed in this work.
In Section~\ref{sect:contour} we provide details on the 
different contour deformation procedures we tested.
In Section~\ref{sect:results} we illustrate the chemical 
potential and volume dependence of the achieved improvement 
and also compare our results with a method based on
Lefschetz thimbles: the holomorphic flow of Ref.~\cite{Alexandru:2015sua}.
We summarize our conclusions in Section~\ref{sect:disc}. 

\section{The chiral random matrix model}
\label{sect:model}
Throughout this paper we will only consider $N_f=2$ with $\mu_1 = \mu_2 \equiv \mu$ for simplicity.
The random matrix model of 
Stephanov~\cite{Stephanov:1996ki} 
for $N_f$ degenerate flavors of quarks is then defined by the partition function
\begin{equation}
\mathcal{Z}_N^{N_f}=e^{N\mu^2}\int\mathrm{d}W\mathrm{d}W^\dagger \left( \mathrm{det}(D+m) \right)^{N_f} e^{-N\mathrm{Tr}WW^\dagger}\rm{,}
\end{equation}
where the massless Dirac matrix is 
\begin{equation}
\label{eq:Dslash}
    D=
        \begin{pmatrix}
            0 & i W+\mu \\
            i W^\dagger+\mu & 0
        \end{pmatrix}\rm{,}
\end{equation} 
$m$ is the quark mass and $W$ is a general $N \times N$ complex matrix. The model has no concept of physical volume. The number of 
degrees of freedom of the model scales with $N^2$. 

The two observables we will study in this paper are the chiral condensate:
\begin{equation}
    \Sigma=\frac{1}{2N}\frac{\partial\log\mathcal{Z}_N^{N_f}}{\partial m}\rm{,}
\end{equation}
and the quark density
\begin{equation}
    n=\frac{1}{2N}\frac{\partial\log\mathcal{Z}_N^{N_f}}{\partial\mu}\rm{.}
    \end{equation}
An important feature of the model is that it can be solved analytically, both in the $N \to \infty$ limit where the integral is dominated by a
saddle point, and at finite $N$ where it reduces to the calculation of moments of Gaussian integrals. Thus, in this particular model
we will be able to compare numerical results with exact analytic solutions.

The model shares with QCD the feature that 
the phase-quenched theory corresponds to an isospin chemical potential, and has an analogue of the 
pion condensation transition at some 
$\mu=\mu^{\mathrm{PQ}}_{c}$. For chemical potentials exceeding $\mu^{\mathrm{PQ}}_c$ the sign problem of the model is severe.
From the point of view of the Dirac spectrum, for 
$\mu = 0$ the eigenvalues are purely imaginary, 
while for $\mu \neq 0$ the eigenvalues of $D$ 
acquire a real part, and
are distributed inside a strip of width $\mu^2$
in the real direction. When the quark mass is 
inside this strip, 
the model has a severe sign problem. This roughly corresponds to the analogue of the pion condensed
phase in the phase-quenched theory.
Due to these similarities, this model has been considered several times in the literature as a good toy model for the 
sign problem in QCD~\cite{Bloch:2017sex,Fukuma:2022yhy}.

We will consider the model for $N_f=2$ and use the same quark chemical potential for both fermion flavors.

In this model, unlike in QCD, the expectation value of the average phase does not always tend to zero in the limit of an infinite system. Rather, it only goes to zero in a given 
range of chemical potentials bounded by the solutions to
the equation~\cite{Han:2008xj}:
\begin{equation}
0=1-\mu^2 + \frac{m^2}{\mu^2-m^2} - \frac{m^2}{4(\mu^2-m^2)^2}\rm{.}
\end{equation}
Using a quark mass of $m=0.2$, the two solutions of this equation are $\mu=0.35=\mu^{\mathrm{PQ}}_{c}$ and $1.02$. This is the regime where the
sign problem in the model is strongest.

\begin{figure*}[t!]
\centering
\includegraphics[width=0.48\textwidth]{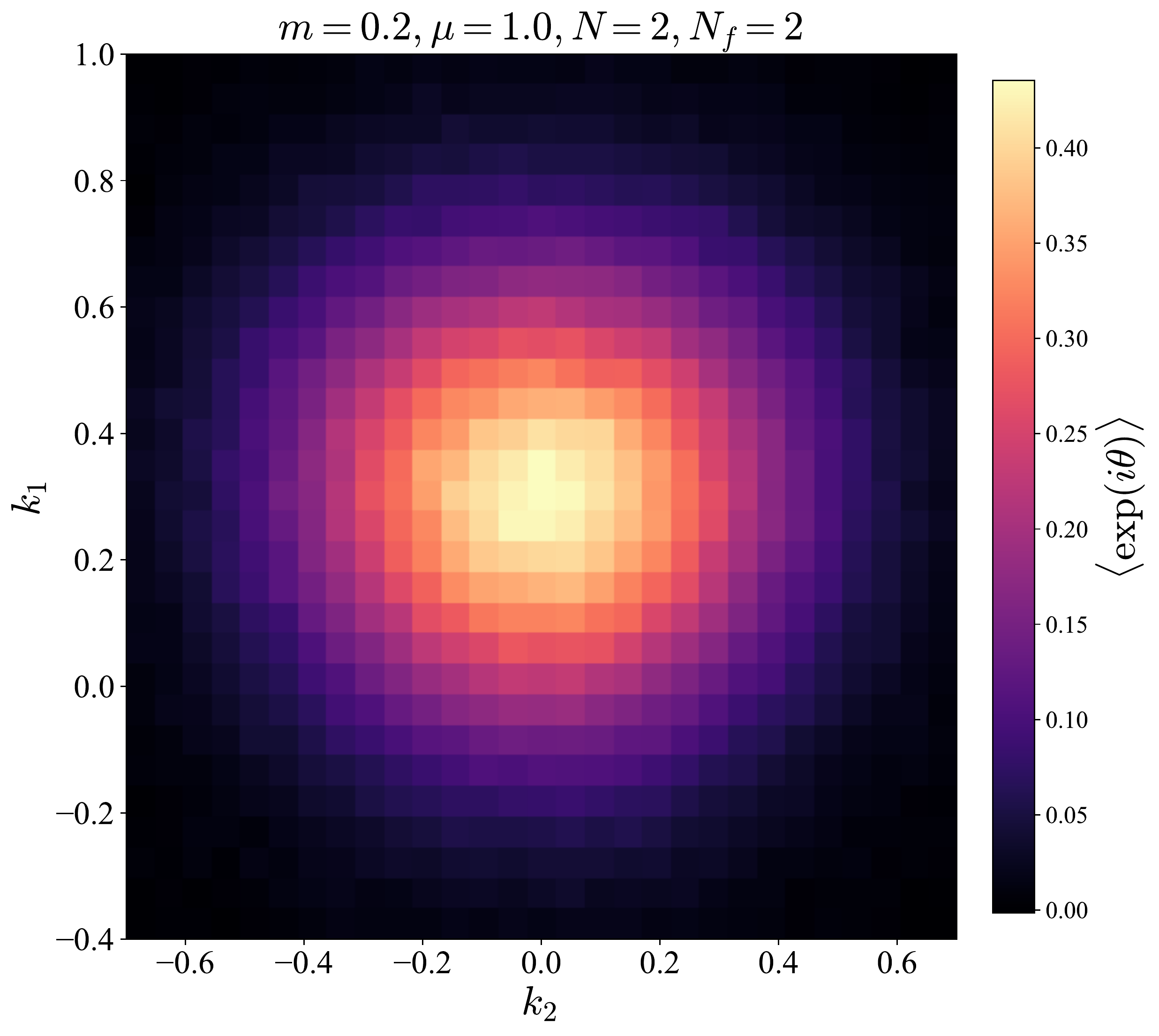}
\includegraphics[width=0.48\textwidth]{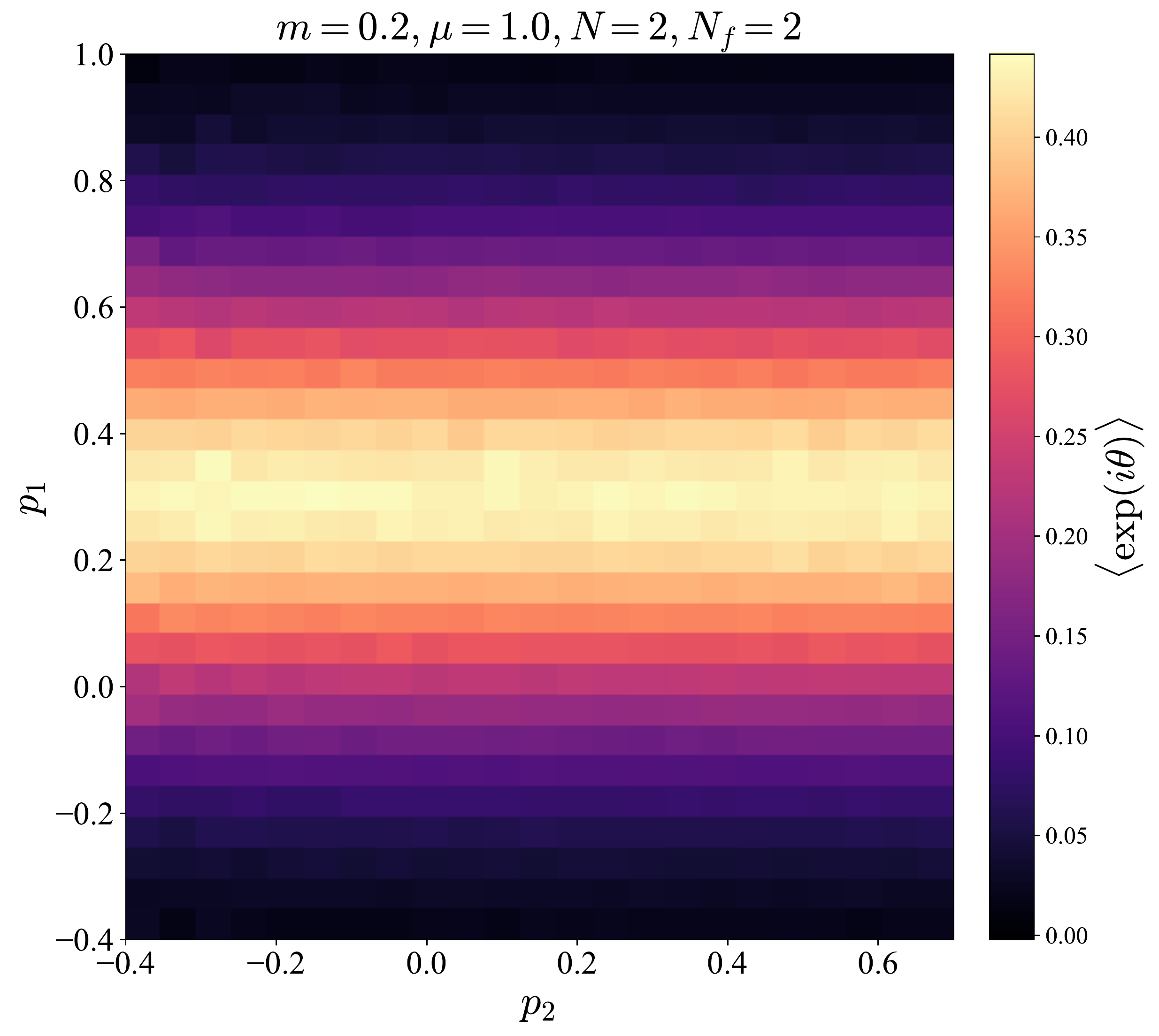}
    \caption{Left: the average phase with Ansatz-1 as a function of $k_1$ and $k_2$. There is a local minimum at $k_2 \approx 0$ and $k_1 > 0$. 
    Right: the average phase with Ansatz-2 as a function of $p_1$ and $p_2$. There is an apparent saddle parallel to the $p_1=0$ line at $p_1=k_1 > 0$.
    }
\label{fig:k1k2_p1p2}
\end{figure*}

\section{Contour deformation methods}
\label{sect:contour}
\subsection{Optimization method}
We will restrict ourselves to ansätze with simple, analytically calculable Jacobians with $\mathcal{O}(N^0)$
computational cost and a small number of parameters, independent of the number of degrees of freedom.

    Let $A=\operatorname{Re} W$ and $B = \operatorname{Im} W$. These two real matrices will be deformed 
    to complex matrices $\alpha$ and $\beta$. Thus, 
    \begin{equation}
        \begin{aligned}
        W &=A+i B \to X=\alpha+i\beta \rm{,}\\ 
        W^\dagger&=A^\mathrm{T}-i B^\mathrm{T}\to Y=\alpha^\mathrm{T}-i\beta^\mathrm{T}\rm{.}
        \end{aligned}
    \end{equation}
    After applying such a deformation $X^\dagger \neq Y$. After the deformation, the severity of the sign problem is given by:
    \begin{equation}
        \langle e^{i\theta}\rangle
        =\bigg\langle\bigg[\frac{\mathrm{det}(D+m)\mathrm{det}\mathcal{J}}{|\mathrm{det}(D+m)\mathrm{det}\mathcal{J}|}\bigg]^{N_f}e^{-iN\mathrm{ImTr}XY}\bigg\rangle\rm{,}
    \end{equation}
    where the Jacobian determinant is
    \begin{equation}
        \mathrm{det}\mathcal{J}=\bigg|\frac{\partial(\alpha,\beta)}{\partial(A,B)}\bigg|\rm{.}
    \end{equation}

\subsection{Holomorphic flow}
Using the holomorphic flow (or generalized thimble method) of Ref.~\cite{Alexandru:2015sua} for the complexified 
action of the Stephanov model,
\begin{equation}
S= -N\mu^2 - N_f \log \mathrm{det}(D+m) + N\mathrm{Tr}(XY) \rm{,}
\end{equation}
we deform the integration manifold by evolving the original one with the differential equation 
\begin{align}
\frac{\drm Y_{ij}}{\drm t} &= \overline{\frac{\partial S}{\partial Y_{ij}}} = N \overline{X}_{ji} - N_f[(\overline{XG})_{ji}+i\mu \overline{G}_{ji}],
\end{align}
where the overbar denotes complex conjugation, $t$ is the flow parameter and 
\begin{equation}
G = \Big[m^2 - \mu^2 - i\mu(X+Y) + YX\Big]^{-1}.
\end{equation}
Solving this system of equations with initial conditions $X_0=W$, $Y_0=W^\dagger$ for a fixed flow time $t_{\mathrm{f}}$ we obtain a deformed manifold $\mathcal{M}_{t_{\mathrm{f}}}$. We parametrize each point on the flowed manifold by the real matrices $A$ and $B$. I.e., we parametrize the flowed manifold
by the initial conditions of the flow equation.

The computation of expectation values requires the Jacobian of the holomorphic flow,
\begin{equation}
\det J = \left|\frac{\partial(X,Y)}{\partial(A,B)}\right|,
\end{equation}
as well. Denoting the Hessian with $H$, the Jacobian matrix $J$ is obtained as the solution of the equation
\begin{equation}
\frac{\drm J}{\drm t} = \overline{H\,J},
\end{equation}
with initial conditions
\begin{equation}
J_{X_{ij},A_{ij}} = 1,\;J_{X_{ij},B_{ij}} = i,\;
J_{Y_{ij},A_{ji}} = 1,\;J_{Y_{ij},B_{ji}} = -i.
\end{equation}
Computing the Jacobian directly is numerically expensive, so we estimate it \cite{Alexandru:2016lsn} with 
\begin{equation}
W = \exp\Bigg[\int_0^T\drm t\;\Tr\,\overline{H(t)}\Bigg].
\end{equation}
The difference between $W$ and $\det J$ is taken into account by reweighting when computing observables,
\begin{equation}
\langle\mathcal{O}\rangle = \frac{\langle\mathcal{O}e^{-\Delta S}\rangle_{S'_{\rm{eff}}}}{\langle e^{-\Delta S}\rangle_{S'_{\rm{eff}}}},
\end{equation}
where $S'_{\rm{eff}} = S - \ln W$, $\Delta S=S_{\rm{eff}}-\mathrm{Re}S'_{\rm{eff}}$ and $\langle.\rangle_{S'_{\rm{eff}}}$ is the average with respect to $e^{-\mathrm{Re}S'_{\rm{eff}}}$. This way, we needed to compute $\det J$ exactly only for the configurations used for measurements.

In the large flow time limit, the flowed manifold tends towards the Lefschetz thimbles. At smaller flow times, it still reduces the sign problem, although less than a complete thimble decomposition would.

\section{Numerical results}
\label{sect:results}

\subsection{Simple ansätze}

     As a rule, all of our ansätze have been parametrized such that the undeformed integration manifold is at value zero for all optimizable parameters. 

\subsubsection*{Ansatz-1}
\begin{figure*}[t!]
\centering
\includegraphics[width=0.48\textwidth]{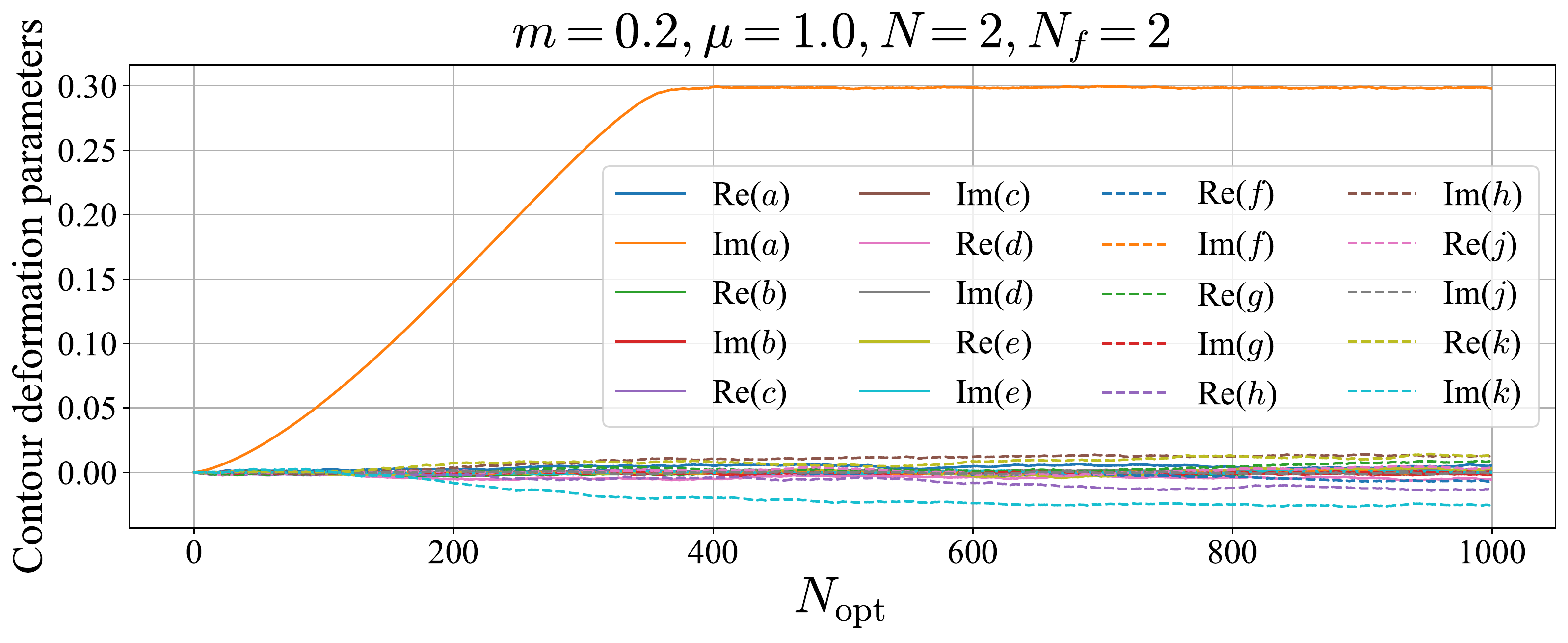}
\includegraphics[width=0.48\textwidth]{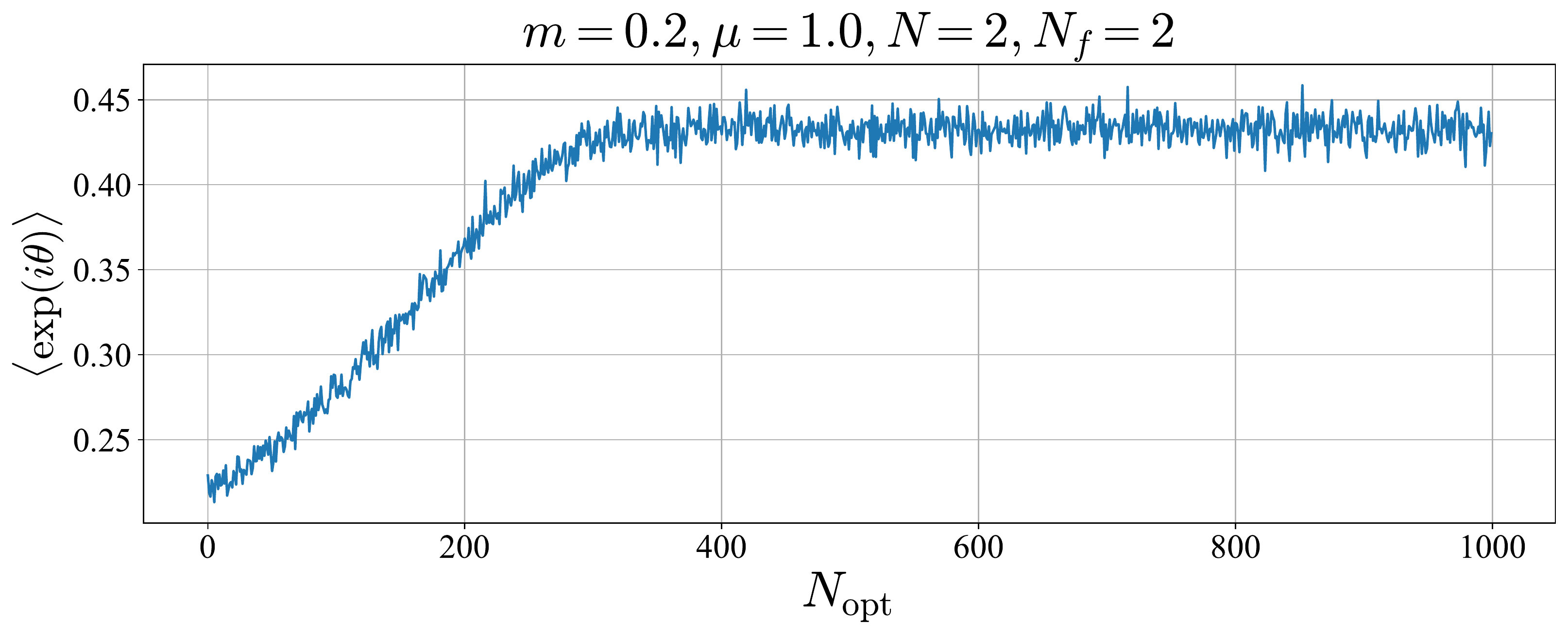}
    \caption{Left: parameters as a function of the optimization step for Ansatz-3.
    Right: the average phase as a function of the optimization step for Ansatz-3.}
\label{fig:20paramOPT}
\end{figure*}

From the definition in Eq.~\eqref{eq:Dslash} it is easy to see that 
the sign problem can be removed from the quark determinant by a 
simple shift of the form $\alpha = A + i \mu \mathbf{1}$. This, however, introduces a sign problem in the Gaussian term $e^{-N \operatorname{Tr}\left(XY\right)}$. By finding 
a trade-off between the two terms, the severity of the sign problem may be optimized. This motivates our first ansatz, with two real parameters $k_1$ and $k_2$ defined by
     \begin{align}
         \alpha&=A + i k_1\mathbf{1}\\
         \beta&=B + i k_2\mathbf{1}\rm{.}
     \end{align}
     The Jacobian determinant for this ansatz is simply unity. The parameter $k_2$ is introduced on a whim, as the matrices $A$ and $B$ do not have
     to be treated symmetrically.
     The results for the average phase in a scan in these two parameters for $N=2$, $m=0.2$ and $\mu=1.0$ is shown
     in Fig.~\ref{fig:k1k2_p1p2} (left).
     While there is a clearly non-zero optimal value for $k_1$, the optimal value of the $k_2$ parameter is near zero.  This remains true for all values of
     the parameters $N,\mu$ and $m$ we simulated.

\subsubsection*{Ansatz-2}

     When we introduce a shift $A\to A+i k\mathbf{1}$ the argument of the Gaussian term changes according to 
    \begin{align}
        \mathrm{Tr}(XY)=\mathrm{Tr}(AA^\mathrm{T}+BB^\mathrm{T})-Nk^2+2 i k\mathrm{Tr}A \rm{.}
    \end{align}
    This motivates our second ansatz, with two real parameters $p_1$ and $p_2$ defined by
     \begin{align}
         \alpha&=A+ i p_1\mathbf{1}+p_2\mathrm{Tr}A\mathbf{1} \rm{,}\\
         \beta&=B \rm{.}
     \end{align}
     The $p_1$ parameter of this ansatz is identical to
     the $k_1$ parameter of the previous ansatz.
     The Jacobian determinant for this ansatz is simply $\mathrm{det}\mathcal{J}=1+Np_2$, i.e., configuration-independent, and can be ignored.
     The results for the average phase in a scan 
     in these two parameters for $N=2$, $m=0.2$ and $\mu=1.0$ can be seen 
     in Fig.~\ref{fig:k1k2_p1p2} (right). 
     While there is a clearly non-zero optimal value for $p_1=k_1$, the $p_2$ parameter only
     appears to move on a saddle.

\subsubsection*{Ansatz-3}

We now move on to a more complicated 
ansatz with 10 complex (or 20 real) parameters $a,b,c,d,e,f,g,h,j,k$ defined by
     \begin{align}
         \alpha&=(a+b\mathrm{Tr}A+c\mathrm{Tr}B)\mathbf{1}+(1+d)A+eB\\
         \beta&=(f+g\mathrm{Tr}A+h\mathrm{Tr}B)\mathbf{1}+jA+(1+k)B
     \end{align}
     The Jacobian determinant for this ansatz is 
     \begin{equation}
         \begin{aligned}
         \mathrm{det}\mathcal{J}=\big((1+d)(1+k)-ej\big)^{N^2-1} \times \\ 
         \big[\big((1+d)+Nb\big)\big((1+k)+Nh\big) \\ 
             -(e+Nc)(j+Ng)\big]\rm{.}
         \end{aligned}
     \end{equation}
The severity of the sign problem was then optimized via the AdaDelta method~\cite{adadelta}, with the 
objective function
    \begin{equation}
        -\log\langle e^{i\theta}\rangle=-\log\frac{\mathcal{Z}}{\mathcal{Z}_\mathrm{PQ}}=-\log\mathcal{Z}+\log\mathcal{Z}_\mathrm{PQ},
    \end{equation}
    where we suppressed the $N$ and $N_f$ indices for the partition function.
    The gradient with right to the deformation parameters is given by
    \begin{equation}
        \nabla\log\mathcal{Z}_\mathrm{PQ}=-\langle \nabla S_\mathrm{eff}^{\mathrm{A}}\rangle \rm{,}
    \end{equation}
    where
    \begin{equation}
        S_\mathrm{eff}^{\mathrm{a}}=N\mathrm{ReTr}XY-N_f\log|\mathrm{det}M|-\log|\mathrm{det}\mathcal{J}|
    \end{equation}
    with gradient
    \begin{equation}
    \label{eq:SeffA}
        \begin{aligned}
            \nabla S_\mathrm{eff}^{\mathrm{a}} = & N\mathrm{ReTr}\big[(\nabla X)Y+X(\nabla Y)\big] \\
                                &-\frac{N_f}{2}\mathrm{Tr}\big[M^{-1}(\nabla M)+\overline{M}^{-1}(\nabla \overline{M})\big] \\
                                &-\mathrm{Re}\bigg[\frac{\nabla\mathrm{det}\mathcal{J}}{\mathrm{det}\mathcal{J}}\bigg].
        \end{aligned}
    \end{equation}
    Note that for Ansatz-3 the Jacobian is independent of the configuration, and the last term can be dropped from Eq.~\eqref{eq:SeffA}.
    For Ansatz-4, to be discused below, the Jacobian will depend 
    on the configuration, and thus the 
    last term is needed.
    An example of such an optimization run is shown in Fig.~\ref{fig:20paramOPT}. As with the previous two ansätze, only a single parameter emerges $k_1=p_1=\Im{a}$.
     
\subsubsection*{Ansatz-4}

\begin{figure}[t!]
\centering
\includegraphics[width=0.48\textwidth]{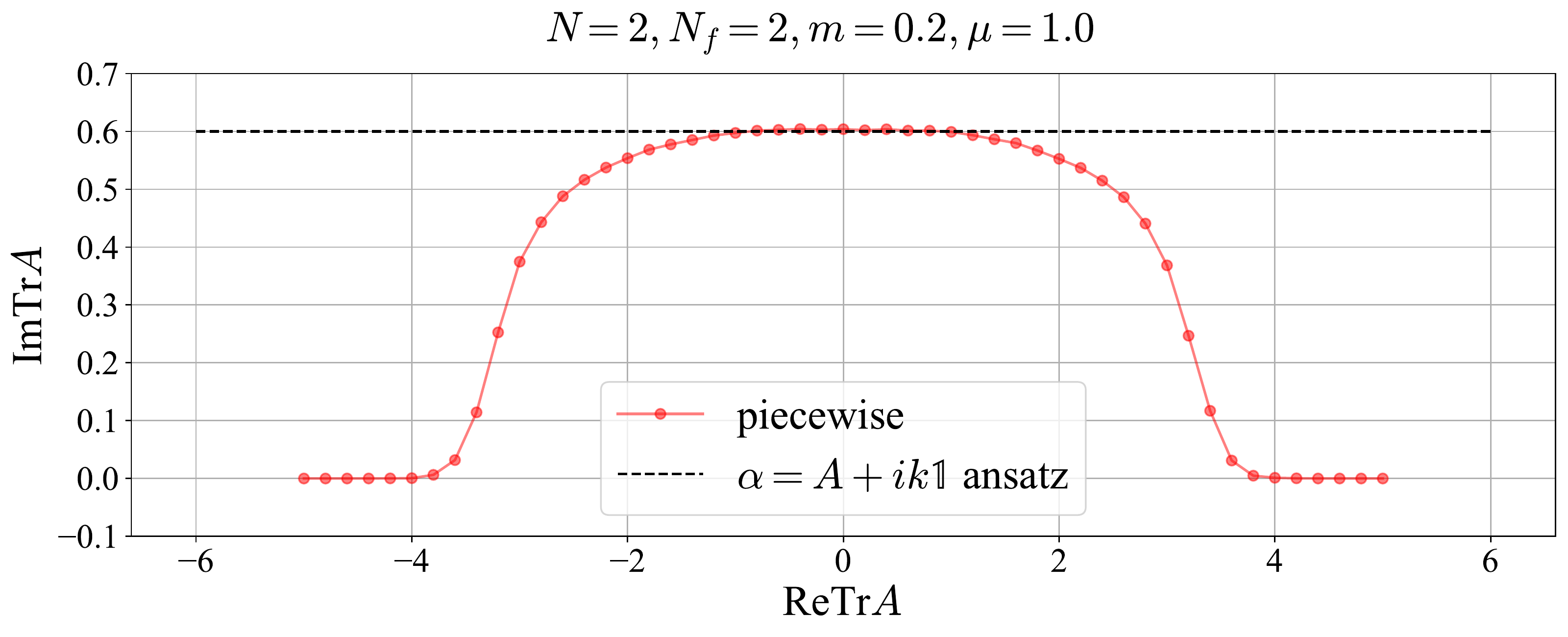}
    \caption{
    Ansatz-4 (piecewise optimization of the trace) compared to Ansatz-1 (imaginary constant shift of $A$ proportional to the unit matrix). The two procedures find essentially the same contour, as the differing tails are at large values of $| \mathrm{Tr} A |$, and have small statistical weight.
    }
\label{fig:piecewise}
\end{figure}

Experiments with the first three ansätze revealed only one parameter of interest, which can be thought of as a simple one-parameter imaginary shift of the trace of the matrix $A$. One might wonder 
whether more general deformations of the trace could lead to a better improvement.
Thus we look at non-linear deformations of the trace $\tau=\mathrm{Tr}A$ of the matrix $A$ with an undeformed $B$ matrix. 
     The integral measure is given by
     \begin{equation}
         \begin{aligned}
         \prod\limits_{i,j=1}^N\mathrm{d}A_{ij}=\mathrm{d} \tau \prod_{\substack{i,j=1\\(i,j)\neq(N,N)}}^{N}\mathrm{d}A_{ij} \\ 
             =\mathrm{d}\tau\prod\limits_{\substack{i,j=1\\i\neq j}}^N\mathrm{d}A_{ij}\prod\limits_{k=1}^N\mathrm{d}\bigg(A_{kk}-\frac{\tau}{N}\bigg)\rm{.}
         \end{aligned}
     \end{equation}
     The deformed matrix $\alpha$ is obtained from $A$ as
     \begin{equation}
         \begin{aligned}
         A=\frac{\tau}{N}\mathbf{1}+\bigg(A-\frac{\tau}{N}\mathbf{1}\bigg)=\frac{\tau}{N}\mathbf{1}+\Tilde{A}\\
         \to\quad\alpha=\frac{\tau}{N}\mathbf{1}+\Tilde{A}\rm{,}
         \end{aligned}
     \end{equation}
     where $\mathrm{Tr}\Tilde{A}=0$ and
     \begin{equation}
         \tau=t+i f(\tau;\dots)\rm{,}
     \end{equation}
     for some function $f$ that depends on $\tau$ and possibly other parameters. For simplicity, we choose $f$ to be piecewise linear, 
     \begin{equation}
     \begin{aligned}
         f(\tau;x_{k(\tau)},x_{k(\tau)+1},y_{k(\tau)},y_{k(\tau)+1})= \\
         \frac{y_{k(\tau)}(x_{k(\tau)+1}-\tau)}{x_{k(\tau)+1}-x_{k(\tau)}}+\frac{y_{k(\tau)+1}(\tau-x_{k(\tau)})}{x_{k(\tau)+1}-x_{k(\tau)}} \rm{.}
     \end{aligned}
     \end{equation}
     The parameters to optimize are the $y_i$, while the node points $x_i$  of the linear interpolation are fixed parameters, and chosen with regular spacing,
     $x_{l+1}-x_l=\Delta$ for all $l$, and 
     \begin{equation}
         k(\tau)=\mathrm{floor}\bigg[\frac{\tau-x_0}{\Delta}\bigg]\rm{.}
     \end{equation}
     By numerical experimentation we have found that the choice of the node points is not important, as long as 
     the full interpolation range is large enough to cover the most probable values of $\mathrm{Tr}A$ on the original contours and $\Delta$ is small enough. If these 
     conditions are met, optimal contours with ans\"atze with different node points appear to be piecewise approximations of the same smooth curve.
     The Jacobian is
     \begin{equation}
         \mathrm{det}\mathcal{J}=1+i\frac{y_{k(\tau)+1}-y_{k(\tau)}}{\Delta}\rm{.}
     \end{equation}   
     The parameters are then optimized as with Ansatz-3.      
     A comparison of the results from this ansatz with
     the constant shift found using ansätze 1 to 3 is shown in Fig.~\ref{fig:piecewise}. For highly 
     probable values of $\operatorname{Tr} A$ the
     two ans\"atze agree, while for the highly improbably
     values of $\operatorname{Tr} A$, the optimization
     does not move the ansatz away from the original
     contour, as there are no configuration to use
     for the optimization of that part of the contour.
     These two asymptotic regimes are 
     smoothly connected. The measured sign problem
     on this contour is identical to the one measured
     with ansätze 1 to 3, up to statistical errors -- not surprisingly since deviations of $f$ from a constant happen on unimportant configurations.

\begin{figure*}[ht]
\centering
\includegraphics[width=0.48\textwidth]{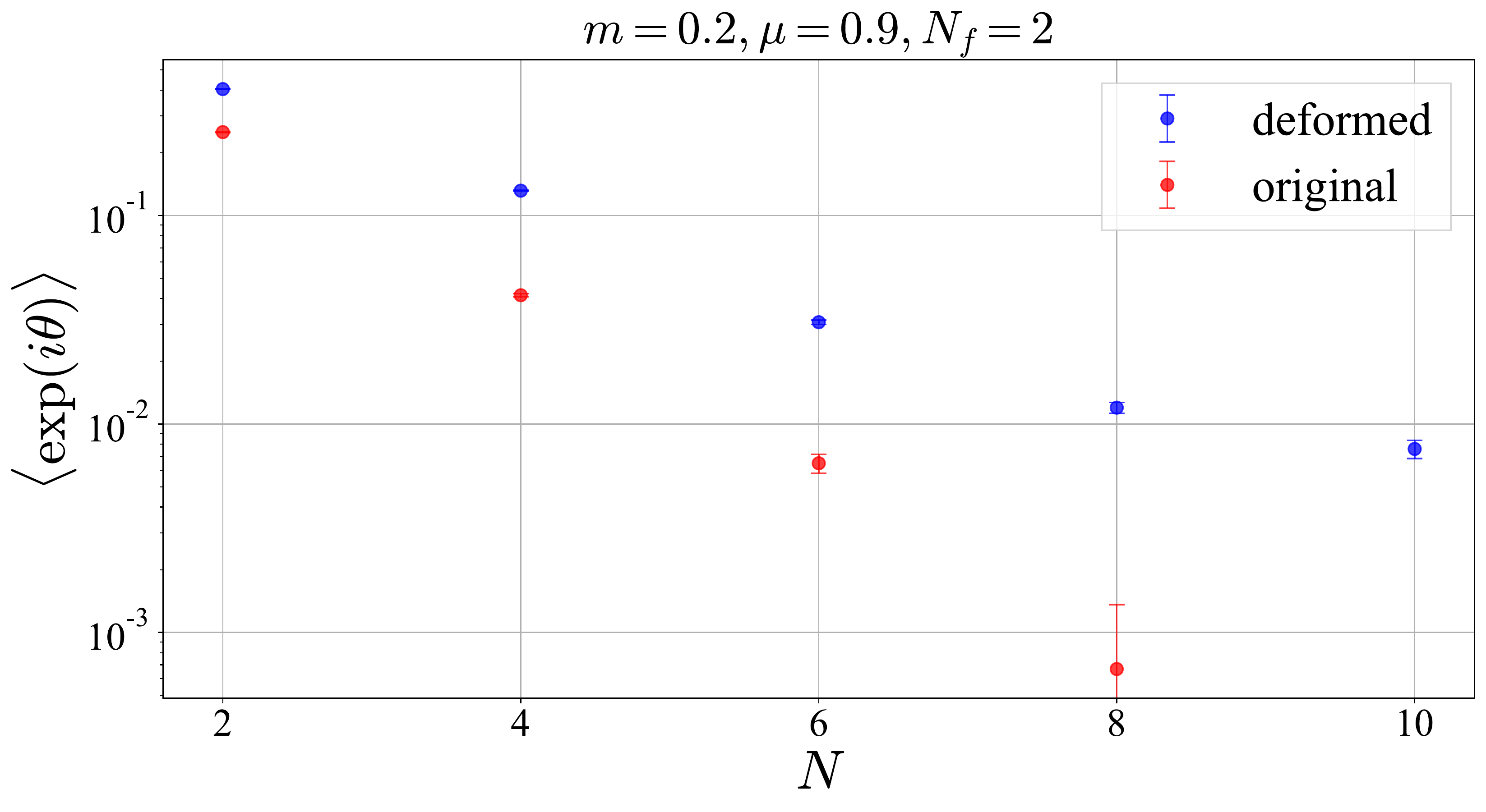}
\includegraphics[width=0.48\textwidth]{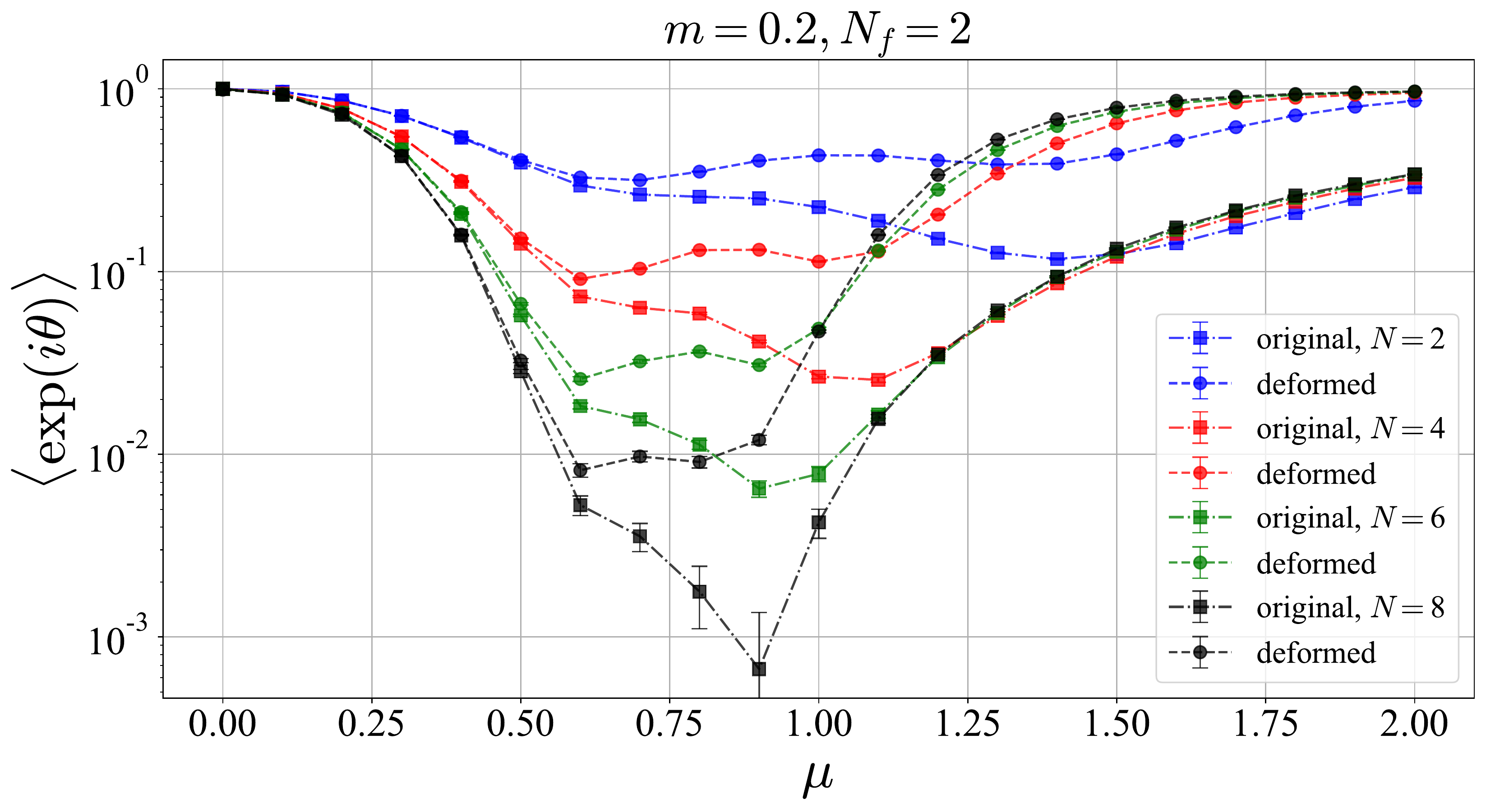}
\caption{Left: dependence of the average phase on the size of the random matrix for the original and optimized contours.
    Right: dependence of the average phase on the chemical potential for the original and optimized contours.}
\label{fig:avphase}
\end{figure*}

\begin{figure*}[ht]
\centering
\includegraphics[width=0.48\textwidth]{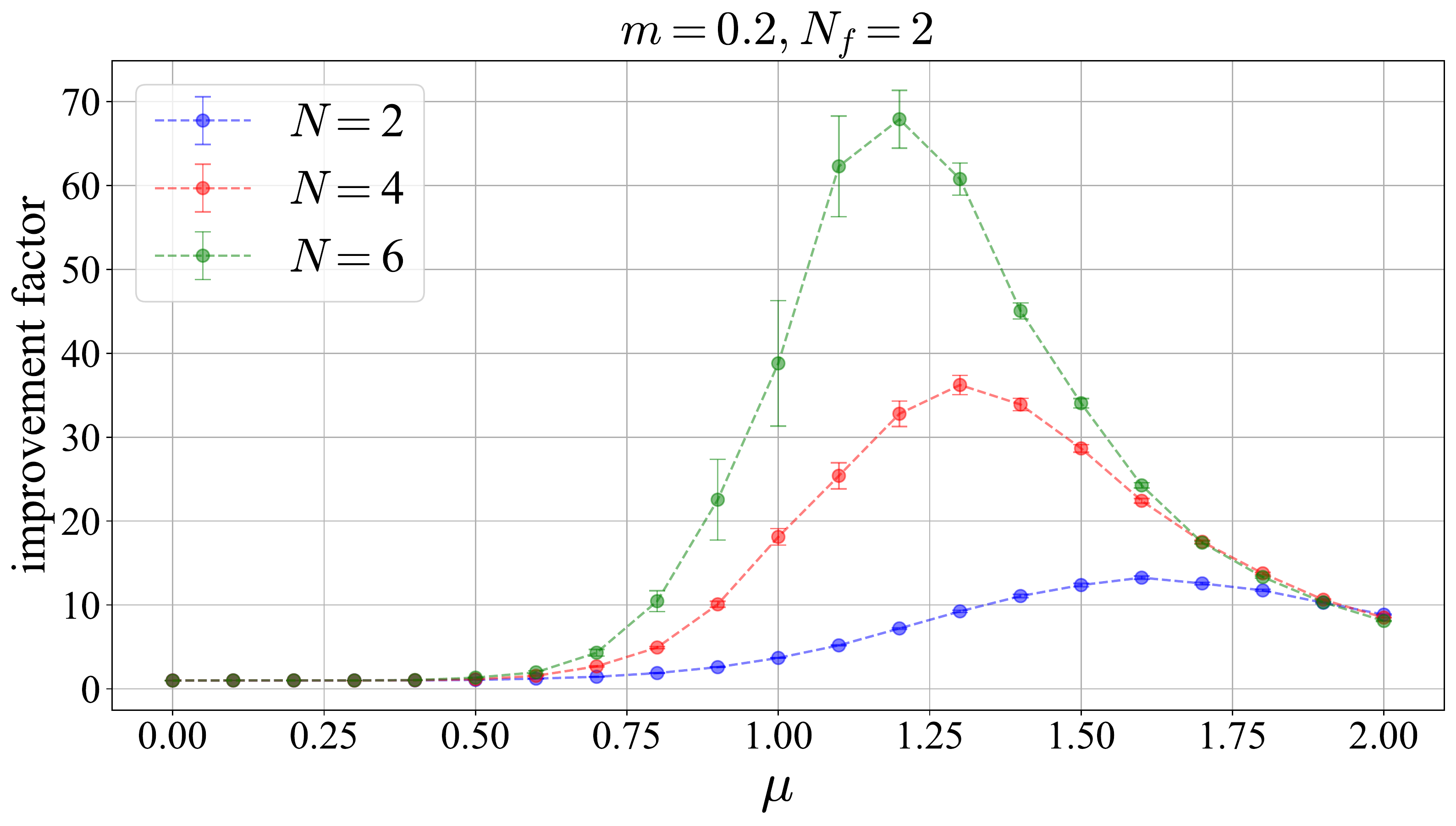}
\includegraphics[width=0.48\textwidth]{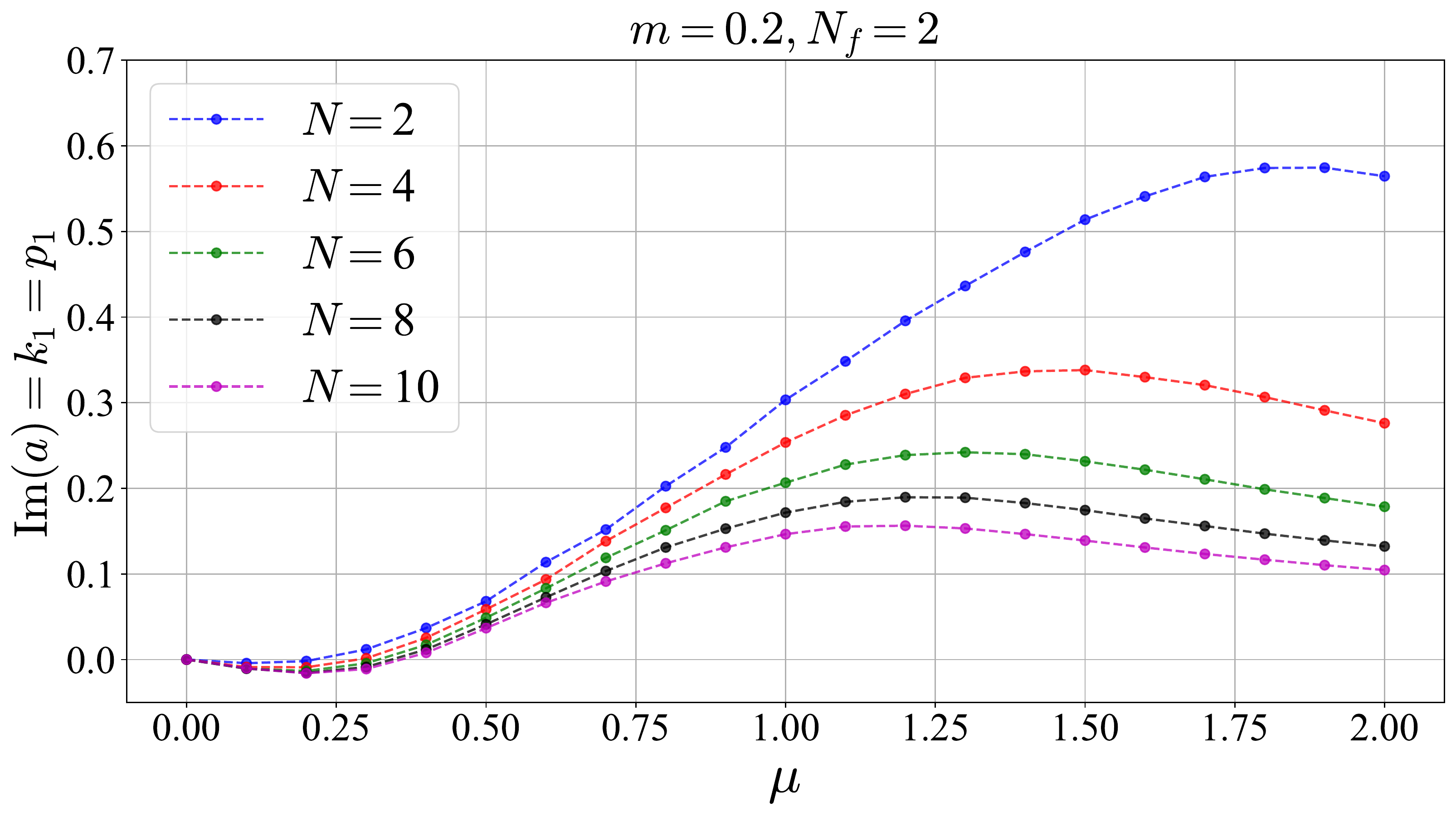}
\label{fig:improvement_and_parameters}
    \caption{Left: dependence of the statistical improvement (calculated as the square of the ratio of the average phases on the optimized and original contours) 
          achieved by contour optimization as a function of $\mu$ for different matrix sizes.
    Right: dependence of the optimal contour parameter $k_1=p_1=\Im a$ on $\mu$ for different matrix sizes.}
\label{fig:improvement_and_parameters}
\end{figure*}

\begin{figure*}[ht]
\centering
\includegraphics[width=0.48\textwidth]{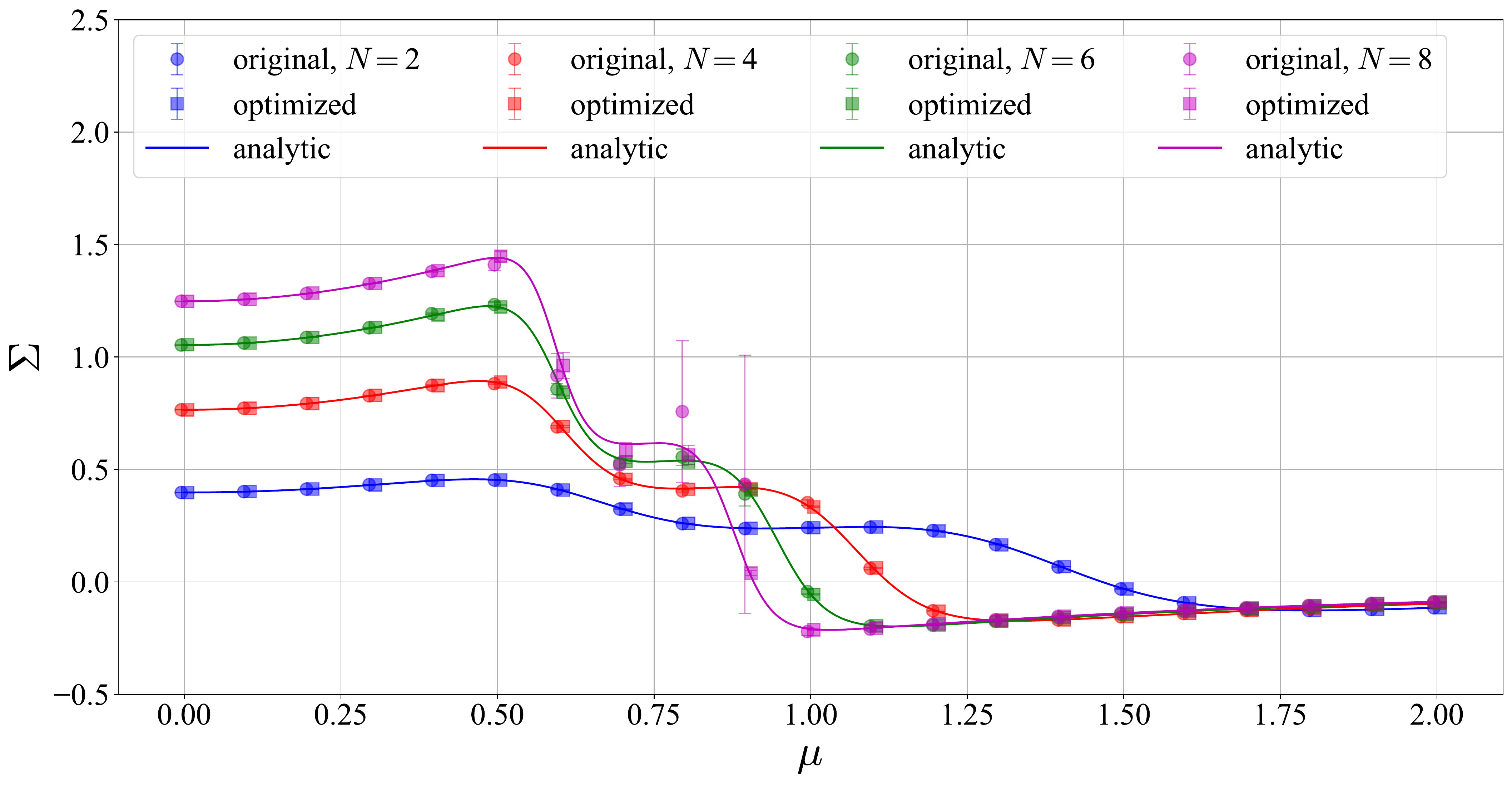}
\includegraphics[width=0.48\textwidth]{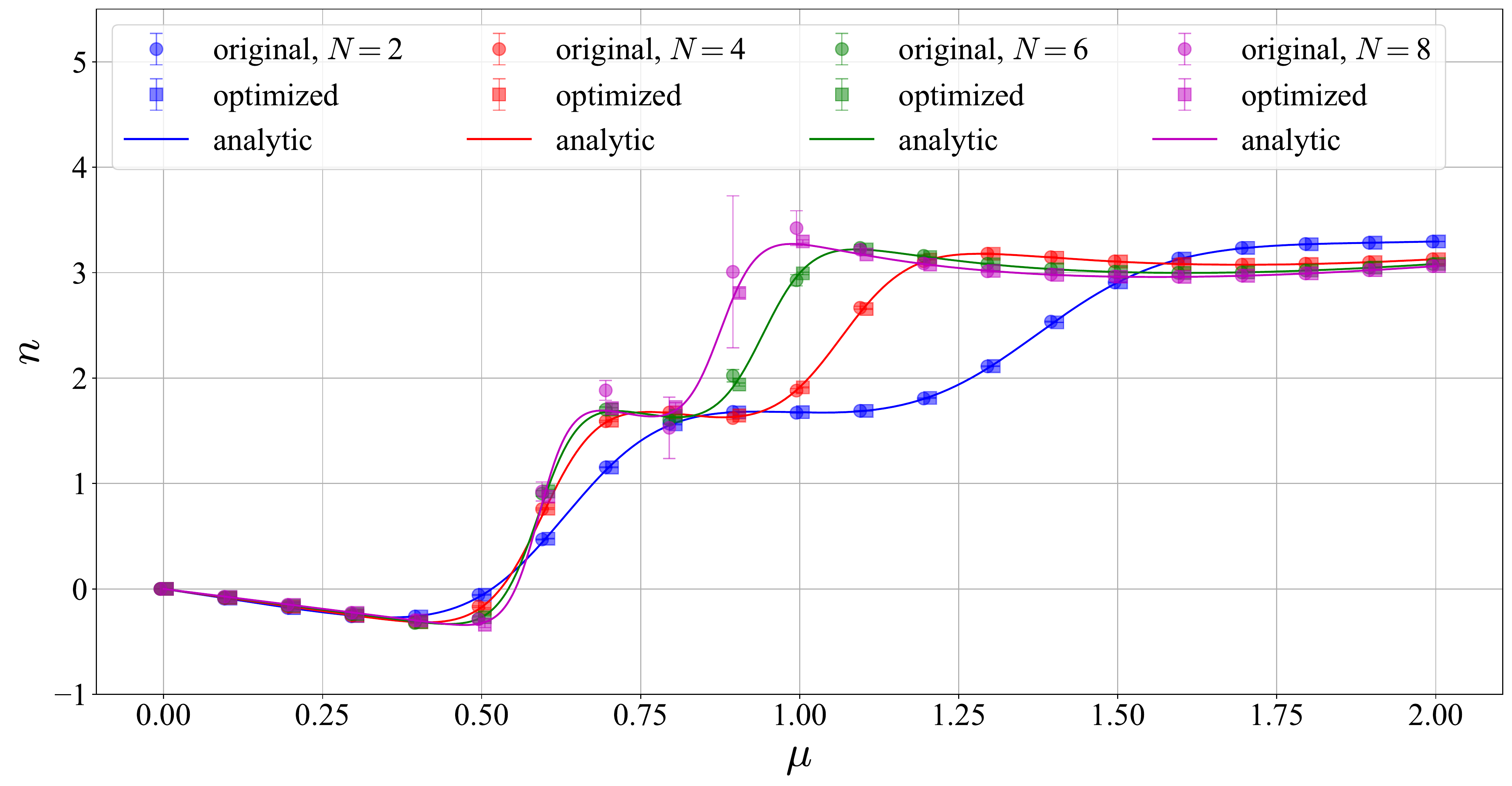}
    \caption{The chiral condensate (left) and the quark number (right) as a function of $\mu$ for 
    several values of the matrix size $N$. Analytic
    results are compared with results from simulations on the original and on the improved
    contours.}
\label{fig:observables}
\end{figure*}

\subsection{Chemical potential and matrix size dependence}

Now that we have discovered a good contour deformation
parameter, let us look at what kind of improvements can
be achieved by such a 1-parameter deformation. From here on out we show 
results with Ansatz-1, with $k_2$ set to zero.

The volume and chemical potential dependence of the average phase for the original and optimized contours is shown in Fig.~\ref{fig:avphase}. The "volume", i.e., matrix size dependence at a fixed chemical potential in the left panel reveals an improvement on the sign problem that is exponential in the matrix size: while the severity 
of the sign problem is roughly linear on a logarithmic
plot for both the original and optimized contours, 
the slopes are quite different. The right panel 
shows the chemical potential dependence for several
values of $N$. Apparently, contour optimization
improves the most on the sign problem in the
regime where it is the most severe. 

The statistical
improvement factor, defined as the square of the ratio of the average phase on the deformed vs the original contours, $\left( \left<e^{i \theta}\right>_{\rm{orig}} / \left<e^{i \theta}\right>_{\rm{def}} \right)^2$, is shown on the left panel of Fig.~\ref{fig:improvement_and_parameters} for $N=2,4$ and $6$. For larger
matrices, $\left<e^{i \theta}\right>$ was zero within statistical errors 
on the original contours,
and this ratio could not be calculated. We see that the ratio monotonically increases with $N$, and as a function of $\mu$ it is maximal close to the value of $\mu$ where the sign
problem is the strongest. The optimal values for the deformation parameter
$k_1=p_1=\Im{a}$ for different values of $\mu$ and $N$ are shown in the right 
panel of Fig.~\ref{fig:improvement_and_parameters}. 

As a sanity check, we also calculated the expectation value of the chiral condensate
and the quark number on both the original and the optimized contours, and compared
them to the analytic results, see Fig.~\ref{fig:observables}. They both show excellent agreement, but the optimized
contours have significantly smaller error bars.

\subsection{Comparison with the holomorphic flow}
\label{sect:flow_compare}

As experiments with simple ansätze so far 
revealed only a single important contour 
deformation parameter, it is a natural question
to ask whether Lefschetz-thimble based methods also
``find'' this deformation or not, and whether 
by utilizing such methods it is possible to 
improve the sign problem further compared 
to such a 1-parameter deformation. For this reason,
we performed the holomorphic flow on our $N=2$ random
matrices, and obtained an estimate of the $k_1$ parameter from the flowed variables via:
$k_1^{\mathrm{flow}} = \operatorname{Im} \left< \operatorname{Tr}(\alpha(t_{\mathrm{f}}) - A) \right>/N$. 
This $k_1$ can then be substituted back 
to the 1-parameter ansatz $\alpha = A + i k_1 \mathbf{1}$ and the severity 
of the sign problem can be compared with the properly flowed manifold. 

\begin{figure}[b!]
\centering
\includegraphics[width=0.48\textwidth]{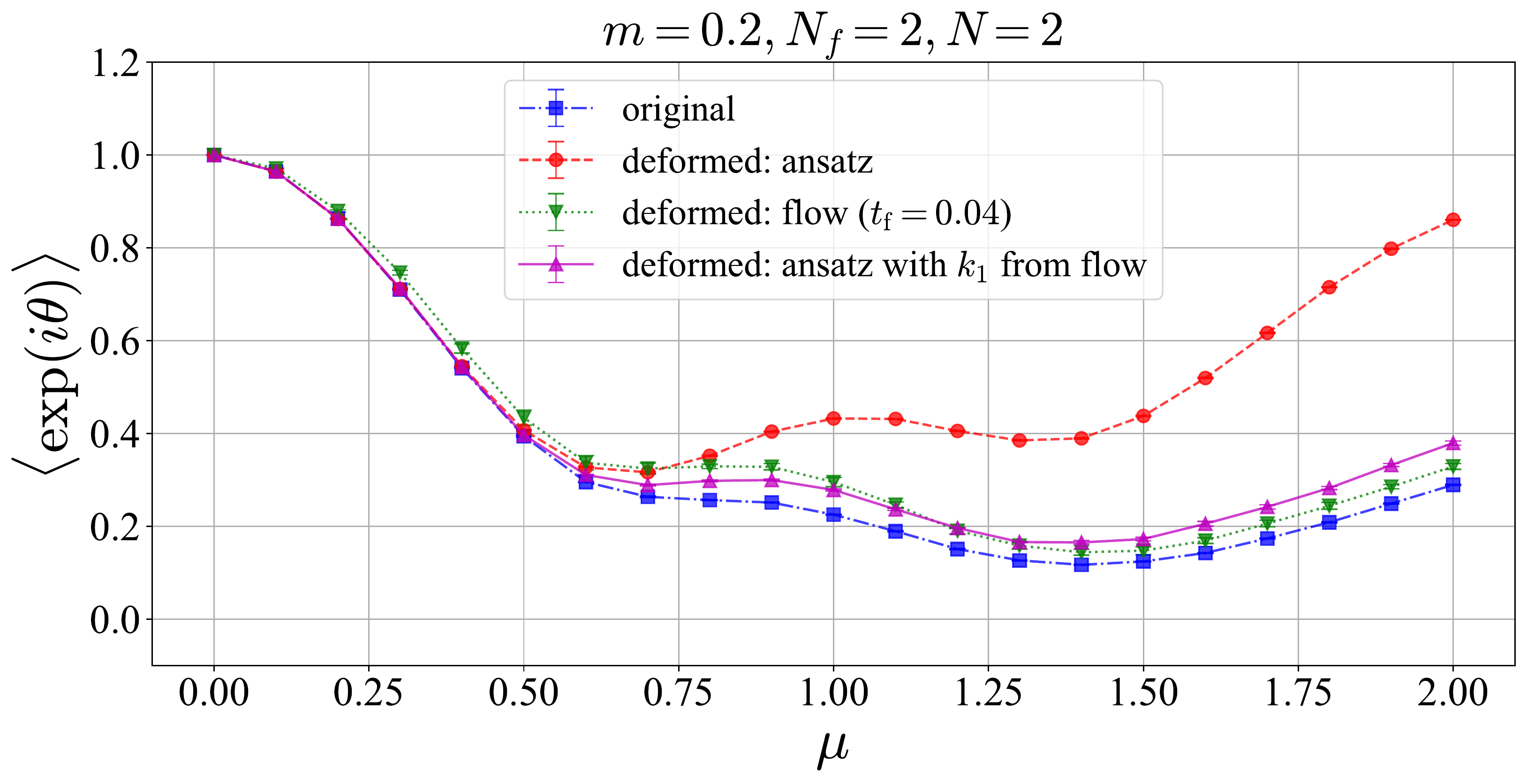}
\caption{\label{fig:compare_flow_k1}
The severity of the sign problem for $N=2$ and $m=0.2$ as a function of $\mu$ on the original contours, the optimized contours, the flowed contours and the contours where the $k_1$ parameter of the 
ansatz is extracted from the flow.}
\end{figure}

The sign problem as a function of $\mu$ is shown on the original contour, the optimized contour, the flowed contour, and on the contour with $k_1$ extracted from the flow in Fig.~\ref{fig:compare_flow_k1}. A few observations can
be drawn from this figure. For small chemical
potentials, the flow performs better than the optimization, which does not noticeably improve the
sign problem. For larger chemical potentials, optimization
vastly outperforms the flow. Of course, this is only 
compared at a fixed flow time, and we do not know
where the severity of the sign problem would end
up at infinite flow time (on the thimbles). However,
going to large flow times gets very 
expensive already for small systems. 

For larger chemical potentials, the 1-parameter
ansatz with $k_1=k_1^{\mathrm{flow}}$ extracted from the flow 
gives very similar results as the full flow. This
may be a hint for the possibility 
that at larger chemical potentials most of the improvement
from the flow comes from this simple deformation.
Interestingly, while the full flow at small chemical
potentials gives a slightly weaker sign problem compared
to the ansatz with $k_1^{\mathrm{flow}}$, at larger chemical potentials the situation is reversed: the sign 
problem is slightly weaker with $k_1^{\mathrm{flow}}$ than
with the solution of the full flow equation. While
this may be somewhat surprising at first, it is not
in contradiction with what we already now 
about contour deformations. The flow goes towards 
the Lefschetz thimbles, which are not the numerically
optimal contours, and thus there is no reason for the
full flow curve to be always above the curve with 
the simple ansatz with $k_1^{\mathrm{flow}}$.

\section{Summary and discussion}
\label{sect:disc}
We have discussed contour deformations in the chiral random matrix model of Stephanov as a way to alleviate its sign problem. Using simple
ad-hoc ansätze we identified a single important
deformation parameter, which allowed for an exponential reduction in the severity of the sign problem as a function of the matrix size.

Our results are quite encouraging, as they show that a simple one-parameter optimization can lead to exponentially alleviating the sign problem even in a fermionic theory, where the thimble decomposition is complicated and contour deformation approaches based on them might not be numerically effective.
The fermionic nature of the matter fields does not appear to be a fundamental obstruction in the construction 
of exponentially better contours. 

Furthermore, the phase diagram of the random matrix model is similar to what we expect in full QCD: the chiral phase transition
is ``hidden behind'' the pion condensation phase in the phase-quenched theory. Hence, this bulk thermodynamic feature -- the existence of 
a phase transition in the phase-quenched theory -- also does not appear to be a fundamental obstruction.

The results and the ansätze in this paper, however, cannot be used directly to construct a good optimization ansatz
in full QCD, as the toy model studied here and QCD differ on an important technical aspect. Concretely, in the Stephanov model there are contour deformations
that can remove the 
sign problem from the fermion determinant for a single flavor (so from the full determinant when all chemical potentials are equal)
 -- albeit at the cost of reintroducing it somewhere else in the 
Boltzmann weights. There are no such deformations in full QCD. The complexification of the $\mathrm{SU}(3)$ gauge group is the $\mathrm{SL}(3,\mathbf{C})$ group, which still requires a unit determinant. To remove the chemical potential
from a single quark determinant the time-like links would have to be deformed to $\mathrm{GL}(3,\mathbf{C})$ matrices, with non-unit 
determinant, which lie outside the complexified gauge group.

Comparison with the holomorphic flow method shows that as one goes near the Lefschetz thimbles in this  model, the bulk (but not all) 
of the improvement on the severity of the sign problem is captured by these types of deformations -- which have no direct analogue in QCD. 
In the future it will therefore be important to work with more realistic toy models of QCD or even full QCD itself,
as the choice of a suitable sign-problem improving ansatz appears to be strongly dependent on the exact symmetries and
exact matter content of a given theory. 

\section*{Acknowledgements}
This work was supported by the NKFIH grant KKP-126769. D.P. is supported by the ÚNKP-22-3 New National Excellence Program of the Ministry for Culture and Innovation from the source of the National Research, Development and Innovation Fund.


\end{document}